\documentclass[11pt]{article}

\usepackage[body={16cm,23cm}]{geometry}
\usepackage{epsfig,amsfonts,amssymb}

\def\Bcal{\mathcal{B}}

\def\Mcal{\mathcal{M}}

\def\Bcal{\mathcal{B}}
\def\Dcal{\mathcal{D}}

\def\Kcal{\mathcal{K}}

\def\Mcal{\mathcal{M}}

\def\g{\gamma}

\def\t{\theta}
\def\vt{\vartheta}
\def\beq{\begin{equation}}
\def\eeq{\end{equation}}
\def\be{\begin{displaymath}}
\def\ee{\end{displaymath}}
\def\bea{\begin{eqnarray}}
\def\eea{\end{eqnarray}}
\def\ov{\overline}
\def\bmat{\left(\begin{array}}

\def\ts{\textstyle}
\def\ds{\displaystyle}

\def\Wp{ \raise.4ex\hbox{\textrm{\Large $\wp$}}}
\def\QQ{\mathcal Q}
\def\f{\Psi}

\def\im{\mathop{\hbox{\rm Im}}\nolimits}

\def\TJ{{\mathsf \Theta}}
\def\HJ{{\mathsf H}}

\def\K{{\mathsf K}}
\def\Q{{\bf Q}}
\def\T{{\bf T}}
\def\P{{\mathcal P}}

\def\eqref#1{(\ref{#1})}
\def\?{(?)\marginpar{|?}}

\renewcommand{\author}[1]{\large\rm #1\\ \bigskip}
\newcommand{\address}[1]{{\normalsize\it #1\\}\bigskip}
\renewcommand{\title}[1]{\bigskip\bigskip\Large\bf #1\bigskip\bigskip\\}
\begin{document}


\vglue .3 cm

\begin{center}
\title{Eight-vertex model and non-stationary Lam\'e equation}

\author{        Vladimir V. Bazhanov\footnote[1]{email:
                {\tt Vladimir.Bazhanov@anu.edu.au}} and
                Vladimir V. Mangazeev\footnote[2]{email:
                {\tt vladimir@maths.anu.edu.au}}}

\address{Department of Theoretical Physics,\\
         Research School of Physical Sciences and Engineering,\\
    Australian National University, Canberra, ACT 0200, Australia.}

\end{center}

\setcounter{footnote}{0}
\vspace{5mm}

\begin{abstract}
We study the ground state eigenvalues of Baxter's
  $\Q$-operator for the
eight-vertex model
in a special case when it describes the off-critical
deformation of the $\Delta=-\frac{1}{2}$ six-vertex model.
We show that these eigenvalues
satisfy a non-stationary Schrodinger
equation with the time-dependent potential given by the
Weierstrass elliptic $\Wp$-function where the modular parameter $\tau$
plays the role of (imaginary) time. In the scaling limit the equation
transforms into a ``non-stationary Mathieu equation'' for
the vacuum eigenvalues of the $\Q$-operators in
the finite-volume massive sine-Gordon model at the super-symmetric
point, which is closely related to the theory of dilute polymers on a cylinder
and the Painlev\'e III equation.
\end{abstract}

\newpage
\section{Introduction and Summary}\label{sect:intro}
The $\Q$-operators introduced by Baxter in his pioneering
paper  \cite{Bax72}  on the eight-vertex model
continue to reveal their exceptional properties in the theory of integrable
quantum systems. These operators play a central role in
the remarkable connection of Conformal Field Theory (CFT)
with the spectral
theory of the Schr\"odinger equation \cite{Vor92} discovered a few
years ago \cite{DT99b}. As shown in \cite{DT99b,BLZsdet},
the vacuum eigenvalues of the $\Q$-operators \cite{BLZ97a} in CFT,
with the central charge $c<1$, can be
identified with the spectral determinants
of certain one-dimensional stationary Schr\"odinger  equations.
Some further developments and applications of this connection can found in the
recent review article \cite{DT03a}.

Apart from a few exceptions,
the vacuum eigenvalues of the $\Q$-operators
(considered as functions of the spectral parameter)
do not generally satisfy any ordinary second order differential equation
themselves. One such exception is the case of the $c=0$ CFT where the
these eigenvalues for particular Virasoro vacuum states
are known to satisfy the Bessel differential
equation \cite{Fen99}. Remarkably, a similar property holds for
the lattice counterparts of these eigenvalues in the
$\Delta=-\frac{1}{2}$ six-vertex model \cite{FSZ01,Str01a} for the chain
of an odd number of sites. In this paper we explain how this property is
generalized for the corresponding cases of the lattice eight-vertex model
and the massive finite volume sine-Gordon model with $N=2$
supersymmetry.

We study the eight-vertex model
on a periodic chain of an odd length, $N=2n+1$, $n=0,1,2,\ldots\infty$.
The eigenvalues of the  transfer matrix of the model, $T(u)$,
satisfy Baxter's famous TQ-equation
\beq
T(u)\,Q(u)=\phi(u-\eta)\,Q(u+2\eta)+\phi(u+\eta)\,Q(u-2\eta),  \label{TQ}
\eeq
where $u$ is the spectral parameter,
\beq
\phi(u)=\vt_1^N(u\,|\,q), \qquad q=e^{{\rm i} \pi \tau},
\qquad \rm{Im}\,\tau>0,\label{phi-def}
\eeq
and $\vt_1(u\,|\,q)$ is the standard theta-function with the periods
$\pi$ and $\pi \tau$ (we follow the notation of \cite{WW}). Here we consider a
special case $\eta=\pi/3$,
where the ground state eigenvalue is known \cite{Bax89,Str01a} to have
a very simple form for all (odd) $N$
\beq
T(u)=\phi(u), \qquad \eta=\frac{\pi}{3}.\label{T-simple}
\eeq
The equation \eqref{TQ} with this eigenvalue, $T(u)$,
has two different solutions \cite{BLZ97a,KLWZ97}, $Q_\pm(u)\equiv
Q_\pm(u,q,n)$, which are entire
functions of the variable $u$ and obey the following periodicity
conditions \cite{Bax72,McCoy2}\footnote{The
factor $(-1)^n$ in \eqref{Qper} and
  \eqref{Qper3} reflects our convention for labeling the
  eigenvalues for different $n$, which will be important in Sec. 3}
\beq
Q_\pm(u+\pi)=\pm (-1)^n Q_\pm(u),
\quad Q_\pm(u+\pi\tau)=q^{-N/2}\ e^{-iNu}\ Q_\mp(u),\quad
Q_\pm(-u)=Q_\pm(u). \label{Qper}
\eeq
The above requirements uniquely determine $Q_\pm(u)$ to within a common
$u$-independent normalization factor. For further references it is convenient
rewrite the functional equation for $Q_\pm(u)$ in the form
\beq
\phi(u)Q_\pm(u)+\phi(u+\frac{2\pi}{3})Q_\pm(u+\frac{2\pi}{3})+
\phi(u+\frac{4\pi}{3})Q_\pm(u+\frac{4\pi}{3})=0.\label{TQ1}
\eeq
We show that the functions
\beq
\f_\pm(u)\equiv\f_\pm(u,q,n)=
\frac{\vt_1^{2n+1}(u\,|\,q)}{\vt_1^n(3u\,|\,q^3)}Q_\pm(u,q,n),\label{fpm}
\eeq
which are meromorphic functions of the variable $u$ for any fixed
values of $q$ and $n$,
satisfy the non-stationary Schr\"odinger equation
\beq
6\,q\frac{\partial}{\partial q}\f(u,q,n)=
\Big\{-\frac{\partial^2}{\partial u^2}+
9\, n\, (n+1)\,\Wp(3u\,|\,q^3)+c(q,n)\Big\}\f(u,q,n),
\label{lame-pde}\eeq
where the modular parameter $\tau$ plays the role of (imaginary) time
and
the time-dependent potential is defined through
the elliptic Weierstrass $\Wp$-function \cite{WW}
(our function $\Wp(v\,|\,e^{i\pi\epsilon})$ has the periods
$\pi$ and $\pi\epsilon$).
The constant $c(q,n)$  appearing in \eqref{lame-pde}
is totally controlled by the normalization of $Q_\pm(u)$ and can be
explicitly determined once this normalization is fixed (see Sec.3 below).

Equation \eqref{lame-pde} is obviously related to the Lam\'e
differential equation
and could be naturally called the ``non-stationary Lam\'e equation''.
To our knowledge this equation\footnote{We are indebted
to Prof. I.M.Krichever for
informing us about the work \cite{EK94}.}
(in fact, a more general
equation) first explicitly appeared in \cite{EK94} as a particular case
of the Knizhnik-Zamolodchikov-Bernard equation \cite{KZ84,Ber88}\  \ for
the one-point
correlation function in the $\textsl{sl(2)}$-WZW-model on the torus.
Here, we will not explore this and other \cite{Ols04}
interesting connections of the
equation \eqref{lame-pde} leaving that for the future.

It is fairly trivial to show that the partial differential equation
\eqref{lame-pde} for the meromorphic functions $\f_\pm(u)$ is
equivalent to the Baxter equation \eqref{TQ1}.
Indeed, every solution of
\eqref{lame-pde}, with the required analytic properties in the variable $u$,
implied by \eqref{Qper} and \eqref{fpm}, 
satisfies eq.\eqref{TQ1}. However, the very existence of
exactly two solutions with these properties is by no means trivial
and reflects some rather special features of eq.\eqref{lame-pde}
discussed below.

Equation \eqref{lame-pde} has been discovered by virtue of
a remarkable {\em polynomial property}
of the eigenvalues $Q_\pm(u)$.
Using a combination of analytical and numerical techniques we have
explicitly solved the equation \eqref{TQ1} for
all values of $n\le 10$. We have found that properly normalized eigenvalues
$Q_\pm(u)$ could always be written as
\beq
Q_\pm(u,q,n)=P_{2n+1}^{(\pm)}\left(\vt_3(\frac{u}{2}\,|\,q^{1/2}),
\vt_4(\frac{u}{2}\,|\,q^{1/2}),\g\right), \qquad \g=
-\left[\frac{\vt_1(\frac{\pi}{3}\,|\,q^{1/2})}
{\vt_2(\frac{\pi}{3}\,|\,q^{1/2})}\right]^2,\label{P-prop}
\eeq
where $P_{2n+1}^{(\pm)}(\alpha,\beta,\g)$ are homogeneous polynomials
of the degree $2n+1$ in the variables
$\alpha$ and $\beta$,  with coefficients being
polynomials in the variable $\g$ with {\em integer
  coefficients}.
Then we considered a class of linear second order partial differential
equations
in two variables $u$ and $q$ where the coefficients of the second order
derivatives are independent of $n$  and all other
coefficients are at most second
degree polynomials in $n$.
Eq.\eqref{lame-pde}
was then found as the only equation in this class
satisfied by \eqref{fpm} with all explicitly calculated
polynomials \eqref{P-prop} with $n\le10$. It turned out that
this equation uniquely defines two and only two such polynomials
\eqref{P-prop} for every value $n=0,1,2,\ldots,\infty$.
It would be interesting to clarify the combinatorial
nature of these polynomials, given that the related $\Delta=-\frac{1}{2}$
  six-vertex model is connected to various important enumeration problems
  \cite{SR2001,BdGN}.

In the scaling limit
\beq
n\to \infty, \qquad q\to 0,\qquad t=8\,n q^{3/2}=\textrm{fixed},\label{scaling}
\eeq
the functions \eqref{fpm} essentially reduce to the ground state eigenvalues
$\QQ_\pm(\t)\equiv\QQ_\pm(\t,t)$ of the ${\bf Q}$-operators  of the
restricted massive sine-Gordon model (at the so-called, super-symmetric point)
on a cylinder of the spatial circumference $R$, where $t=M R$ and $M$ is the
soliton mass.  The equations \eqref{TQ} and \eqref{Qper}
become
\beq
\QQ_\pm(\t)=\QQ_\pm(\t+2\pi i)+\QQ_\pm(\t-2\pi i),
\eeq
\beq
\QQ_\pm(\t+3\pi i)=e^{\pm \frac{i \pi}{2}}\,
\QQ_\pm(\t),\qquad \QQ_+(\t)=\QQ_-(-\t),
\eeq
where the variable $\t$ is defined as $u=\pi\tau/2-i\t/3$.
With a suitable $t$-dependent normalization of $\QQ_\pm(\t)$
equation \eqref{lame-pde} could be brought to a particularly simple form
\beq
t\frac{\partial}{\partial t}\
\QQ_\pm(\t,t)=\Big\{\frac{\partial^2}{\partial \t^2}-
{\frac{1}{8}}\,t^2\, (\cosh 2\t-1) \Big\}\ \QQ_\pm(\t,t).
\label{mathieu-pde}
\eeq
With the same normalization the
asymptotic behavior of $\QQ_\pm(\t)$ at large
$\t$ is given by
\bea
\log\, \QQ_\pm(\t)&=&-\frac{1}{4}\,t\, e^\t+\log\Dcal_\pm(t)
+2\left(\partial_t\log \Dcal_\pm(t)-t/8\right)\,e^{-\t}\nonumber\\
&&\ \ +2\left(\partial^2_t\log \Dcal_\pm(t)-\partial_t\log
\Dcal_\pm(t)/t\right)
\,e^{-2\t}
+O(e^{-3\t}),\qquad \t\to+\infty,\label{asymp}
\eea
where $\Dcal_\pm(t)$ are the Fredholm determinants which previously appeared
in connection with the calculation of the ``supersymmetric index''
and the problem of dilute polymers on a cylinder \cite{FS92,FS94,Zam94,TW96}.
Note, in particular,
that the quantity
\beq
F(t)=\frac{d}{dt}\,\,U(t), \qquad U(t)=\log\,\frac{\Dcal_+(t)}
{\Dcal_-(t)}\ ,
\eeq
describes the free energy of a single incontractible polymer loop and
satisfies the Painlev\'e III equation \cite{FS92}
\beq
\frac{1}{t}\frac{d}{dt}t\frac{d}{dt}\, U(t)=\frac{1}{2}\,\sinh 2 U(t).
\eeq
In this connection, it is useful to 
mention the other celebrated appearances of  
Painlev\'e transcendents in the theory of the two-dimensional Ising
model \cite{BMW} and in the problem of isomonodromic transformations 
of the second order differential equations \cite{JMSM}.

We did not attempt to make this paper self-contained. Detailed
results are presented in \cite{BMfuture}. The main
reference for the eight-vertex model and its commuting
$\T$- and $\Q$-matrices is
Baxter's original paper
\cite{Bax72}; Sec.2 is meant to be read in conjunction with this
paper. The definitions of the  $\T$- and $\Q$-operators in continuous
quantum field theory are given in \cite{BLZ97a,BLZ97b}. The connection of the 
$\Q$-operators with the problem of 
dilute polymers is explained in \cite{Fen99}.

\section{The eight-vertex model and TQ-relation}\label{eight}

We consider the eight-vertex model on the $N$-column
square lattice with periodic
boundary conditions and assume that $N$ is an odd integer $N=2n+1$.
Following \cite{Bax72} we parameterize the Boltzmann weights
$a$, $b$, $c$, $d$ of the eight-vertex model as\footnote{We
use the notation of \cite{WW} for theta-functions
  $\vt_k(u\,|\,q)$, $k=1,2,3,4$, of
the periods $\pi$ and $\pi\tau$, $q=e^{i\pi\tau}$, $\im \tau>0$.
The theta-functions $\HJ(v)$, $\TJ(v)$ of the nome $q_B$
used in \cite{Bax72} are given by
$$
q_B=q^2,\quad \HJ(v)=\vt_1(\frac{\pi v}{2\K_B}\,|\,\,q^2),\quad \TJ(v)=
\vt_4(\frac{\pi v}{2\K_B}\,|\,\,q^2),
$$
where  $\K_B$ is the complete elliptic integral of the first kind
with the nome $q_B$.}
\bea
&a=\rho
  \ \vt_4(2\eta\,|\,q^2)\ \vt_4(u-\eta\,|\,q^2)\ \vt_1(u+\eta\,|\,q^2),&
\nonumber\\
&b=\rho
  \ \vt_4(2\eta\,|\,q^2)\ \vt_1(u-\eta\,|\,q^2)\ \vt_4(u+\eta\,|\,q^2),
&\nonumber\\
&c=\rho
  \ \vt_1(2\eta\,|\,q^2)\ \vt_4(u-\eta\,|\,q^2)\ \vt_4(u+\eta\,|\,q^2),
&\label{weights}\\
&d=\rho
  \ \vt_1(2\eta\,|\,q^2)\ \vt_1(u-\eta\,|\,q^2)\ \vt_1(u+\eta\,|\,q^2),
&\nonumber
\eea
and fix the normalization factor $\rho$ as
\beq
\rho=2\,\ \vt_2(0\,|\,q)^{-1}\ \vt_4(0\,|\,q^2)^{-1}.\label{ei2}
\eeq
It is convenient to introduce new variables $\g$ and $x$, 
where $\g$ is defined by
(cf. \cite{Bax72}),
\beq
\g=\frac{(a-b+c-d)(a-b-c+d)}{(a+b+c+d)(a+b-c-d)}=
-\left[\frac{\vt_1(\eta\,|\,q^{1/2})}
{\vt_2(\eta\,|\,q^{1/2})}\right]^2,\label{gamma}
\eeq
and $x$ satisfies the quadratic equation
\beq
\Big(\sqrt{x}-\frac{\g}{\sqrt{x}}\Big)^2=-\frac{16\,(a-b)^2\,c\,d}{(c+d)^2\,
(a+b+c+d)(a+b-c-d)},\label{x-def}
\eeq
where we choose the following root
\beq
x=\g\frac{\ov\vt_3^{\>2}(u)}{\ov\vt_4^{\>2}(u)},
\qquad
\ov \vt_3(u)=\vt_3(\frac{u}{2}\,|\,q^{1/2}),
\quad \ov\vt_4(u)=\vt_4(\frac{u}{2}\,|\,q^{1/2}).\label{xz-def}
\eeq

The row-to-row transfer-matrices ${\bf T}(u)$ and the $\Q(u)$-matrices
of the model form a commutative family;
their eigenvalues satisfy the functional relation \eqref{TQ}.
In \cite{Bax72} Baxter explicitly constructed the
matrix ${\bf Q}(u)$ provided that
$\eta =\frac{\pi m}{2L}$ for integer $m$ and $L$.
We only consider a special case when the weights
\eqref{weights} satisfy the constraint \cite{Str01a}
\beq
(a^2+ab)\,(b^2+ab)=(c^2+ab)\,(d^2+ab),
\eeq
which is equivalent to the condition $\eta=\pi/3$.
In \cite{Str01a,Str01c} it
was conjectured that the largest eigenvalue of the
transfer matrix (corresponding to the double-degenerate
ground state of the model) has the simple form, \eqref{T-simple},
\beq
T(u)=(a+b)^N=\phi(u).\label{tab}
\eeq
Here we study the corresponding eigenvalues $Q_\pm(u)$ of the $\Q$-matrix.
As noted in \cite{McCoy1}, the method used in \cite{Bax72} for the
construction of the $\Q$-matrix cannot be executed in its full strength
for $\eta=\pi/3$, since some axillary $\Q$-matrix, $\Q_R(u)$, in
\cite{Bax72} is not invertible in the full $2^N$-dimensional space of
states of the
model. The numerical results presented in Table~1 of \cite{McCoy1} 
make it natural to 
suggest that the rank of $\Q_R(u)$ in this case 
is given by the $N$-th Lucas number 
$((1+\sqrt{5})/2)^N+((1-\sqrt{5})/2)^N$ which coincides
with the dimension of the space of states of the $N$-site hard hexagon model
\cite{Bax80}\footnote{S.M.Sergeev noted \cite{Sergeev} that 
only a minor modification of the arguments of \cite{Bax72} leads to the matrix 
$\Q_R$ with the rank equal to $2^N$ for even $N$ and to 
$(2^N-2)$ for odd $N$.}. 
This indicates that for $\eta=\pi/3$ the construction of \cite{Bax72} 
only provides 
a ``restricted'' $\Q^{(r)}(u)$-matrix which acts in
some  ``RSOS-projected'' subspace
of the full space of states of
the eight-vertex model. This interesting phenomenon
certainly deserves special investigations in its own right; we have
verified for several small values of $N$ that the ground state
eigenvectors corresponding to \eqref{tab} belongs to this RSOS-projected
subspace and that the eigenvalues of $\Q^{(r)}(u)$ satisfy \eqref{TQ}.
Below we will assume that this is true for all (odd) $N$.

After this brief review, let us now describe our main results.
For $\eta=\pi/3$ the variable $\g\equiv\g(q)$, defined by \eqref{gamma},
depends on $q$ only, while the variable $x\equiv x(u,q)$, 
defined by \eqref{x-def}, depends on $u$ and $q$.
Below it will be  more convenient to use the
combinations
\beq
Q_1(u)=(Q_+(u)+Q_-(u))/2,\quad Q_2(u)=
\,(Q_+(u)-Q_-(u))/2. \label{Q12-def}
\eeq
which are simply related by the periodicity relation
\beq
Q_{1,2}(u+\pi)=(-1)^n\,Q_{2,1}(u).\label{Qper3}
\eeq
Bearing in mind this simple relation we will only quote
results for $Q_1(u)$, writing it as $Q^{(n)}_1(u)$ to indicate the
$n$-dependence.  We have found that all
the eigenvalues $Q^{(n)}_1(u)$ can be written as
\beq
Q^{(n)}_1(u)={\mathcal N}(q,n)\, \ov\vt_3(u)\>\ov\vt_4^{\>2n}(u)\>
\P_n(x,z), \qquad z=\g^{-2},\label{Q1}
\eeq
where ${\mathcal N}(q,n)$ is an arbitrary normalization factor and
$\P_n(x,z)$ are polynomials in $x,z$ of the degree $n$ in $x$,
\beq
\P_n(x,z)=\sum_{k=0}^n r^{(n)}_k(z)\,x^k.\label{P-def}
\eeq
while  $r^{(n)}_i(z)$, $i=0,\ldots,n$,
are polynomials in $z$ with integer coefficients. The normalization of
$\P_n(x,z)$ is  fixed by the requirement $r^{(n)}_n(0)=1$.
The polynomials $\P_n(x,z)$ are uniquely determined by the following
partial differential equation in the variables $x$ and $z$,
\beq
\Big\{A(x,z)\,\partial_x^2 +B_n(x,z)\,\partial_x +C_n(x,z)\, +
T(x,z)\,{\ds\partial_z}\Big\}\ \P_n(x,z)=0,\label{P-pde}
\eeq
where
\bea
A(x,z)&=&2x(1 + x - 3xz + x^2z)(x + 4z - 6xz -3xz^2 + 4x^2z^2),\\
B_n(x,z)&=&4(1+x-3xz+x^2z)(x+3z-7xz+3x^2z^2)+\label{gen8}\\
&&\kern-1.0em
+2nx(1-14z+21z^2-8x^3z^3+3x^2z(3z^2+6z-1)-x(1-9z+23z^2+9z^3))\nonumber\\
C_n(x,z)&=&n\big[z(9z-5)+x^2z(3z^2+11z-2)+x(9z^3-38z^2+19z-2)-\nonumber\\
&&\kern-1.0em
-4x^3z^3+nz(1-9z-x(9z^2-36z+3)+x^2(3z^2-31z+4)+8x^3z^2)\big]\label{gen9}\\
T(x,z)&=&-2z(1-z)(1-9z)(1 + x - 3xz + x^2z).\label{gen10}
\eea
The first few polynomials $\P_n(x,z)$ read
\beq
\P_0(x,z)=1, \quad \P_1(x,z)=x+3,\quad
\P_2(x,z)=x^2(1+z)+5x(1+3z)+10,\label{gen4}
\eeq
\beq
\P_3(x,z)=x^3(1+3z+4z^2)+7x^2(1+5z+18z^2)+7x(3+19z+18z^2)+35+21z.\label{gen5}
\eeq
In the next Section, we will prove that the eigenvalues $Q^{(n)}_1(u)$ given by
\eqref{Q1} (as well as all related eigenvalues $Q^{(n)}_2(u)$ and
$Q^{(n)}_\pm(u)$ given by \eqref{Q12-def} and \eqref{Qper3})
automatically satisfy the functional relation \eqref{TQ1},
merely as a consequence of the defining property
\eqref{P-pde} of $\P_n(x,z)$. Of course, for small
values of $n$ this
functional relation can be checked directly. For example, it is
not very difficult to check it for
\beq
Q^{(0)}_1(u)\sim\ov\vt_3(u),\qquad
Q^{(1)}_1(u)\sim\ov\vt_3(u)\left[\g\>\ov\vt_3^{\,2}(u)
+3\>\ov\t_4^{\,2}(u)\right],
\eeq
by employing various identities for elliptic functions, however for
$n\ge2$ this does not appear to be practical.

The polynomials $\P_n(x,z)$ can be effectively calculated with the
following procedure.  It is easy to see that \eqref{P-pde} leads  to
descending recurrence relations for coefficients in \eqref{P-def} in
the sense that each coefficient $r^{(n)}_k(z)$ with $k<n$ can be
recursively calculated in terms of $r^{(n)}_m(z)$, with $m=k+1,\ldots,n$
and, therefore, can be eventually expressed through the coefficient
$s_n(z)\equiv r^{(n)}_n(z)$ of the leading power of $x$.
These leading coefficients,  $s_n(z)$, $n=0,1,2\ldots\infty$,
are uniquely determined by the following
recurrence relation
\bea
&2z(z-1)(9z-1)^2[\log s_n(z)]''_z+2(3z-1)^2(9z-1)[\log s_n(z)]'_z+&\label{gen19}\\
&\>\>\>+8(2n+1)^2{\ds\frac{s_{n+1}(z)s_{n-1}(z)}{s_n^2(z)}}-
[4(3n+1)(3n+2)+(9z-1)n(5n+3)]=0,\nonumber
\eea
with the initial condition $s_0(z)=s_1(z)\equiv 1$. In particular,
for $z=1/9$ (corresponding to $q=0$) this gives
\beq
 s_0({\ts\frac{1}{9}})=1,\qquad
 s_{n+1}({\ts\frac{1}{9}})=
\frac{2^{n-1}n!(3n+2)!}{3^n((2n+1)!)^2}\ s_n({\ts\frac{1}{9}}).\label{gen20}
\eeq
Currently, the polynomials $\P_n(x,z)$ have been calculated explicitly
for $n\le50$; using the above procedure they can be easily calculated
for higher values of $n$.

\section{Non-stationary Lam\'e equation}

Evidently, the algebraic form of the partial differential equation given by
\eqref{P-pde} is not very illuminating (even though it is
quite useful for the analysis of polynomial solutions).
Fortunately this equation has a much more elegant equivalent form given by the
non-stationary Lam\'e equation \eqref{lame-pde} discussed in the Introduction.
The details of transformations between these two forms  will be presented
elsewhere \cite{BMfuture}. Noting the frustrating expressions for the
coefficients in \eqref{P-pde} it is not surprising that these transformations
turn out to be rather tedious. They involve many elliptic
function identities, in particular, the algebraic properties of the
ring of theta-constants and their $q$-derivatives \cite{Zud00} happened to
be extremely useful.

Let us choose the normalization factor ${\mathcal N}(q,n)$ in
\eqref{Q1} as
\beq
(-1)^n\,2^{(3n^2+5n+1)/2}\,s_n({\ts\frac{1}{9}})\ {\mathcal N}(q,n)=
i^n \, \g^{-n}\,\vt'_1(0\,|\,q)^{-n-1/3}\,
\Big\{(\g^2-1)\,\vt_4^4(0\,|\,q)\Big\}^{\frac{n(n+1)}{2}}
\eeq
and define the functions $\f_\pm(u,q,n)$ as in \eqref{fpm} where
$Q_\pm(u)$ are given by \eqref{Q12-def}, \eqref{Qper3} and \eqref{Q1}.
Then the equation \eqref{P-pde} takes the form \eqref{lame-pde} where the
constant term $c(q,n)$ is given by
\beq
c(q,n)=-3\,n\,(n+1)\,\frac{\vt_1'''(0|q^3)}{\vt_1'(0|q^3)}.
\eeq
By construction the functions $\f_\pm(u,q,n)$ are meromorphic
functions of the variable $u$, which obey periodicity relations
\beq
\f(u+2\pi)=\f(u),\qquad
\f(u+2\pi\tau)=q^{-6}\,e^{-6\/iu}\ \f(u),\qquad\f(-u)=(-1)^{n+1}\f(u)
\label{f-per}
\eeq
and have $(n+1)$-th order zeros at $u=k\pi+m\pi\tau$, where
$k,m\in{\mathbb Z}$, i.e., 
\beq
\f(\varepsilon+k\pi+m\pi\tau)=O(\varepsilon^{n+1}),\qquad 
\varepsilon\to0, \qquad k,m\in{\mathbb Z}.\label{psi-cond}
\eeq
Let us show that for $n\ge1$,
equation \eqref{lame-pde}, restricted to a class of functions $\f(u)$
with such analytic properties, is
equivalent to the functional relation \eqref{TQ1}.
Any such solution  of \eqref{lame-pde}, $\f(u)$, could have either
an $(n+1)$-th order zero or an $n$-th order pole in the variable $u$,
at all points $u=(3k\pm1)\pi/3+m\pi\tau$, with  $k,m\in{\mathbb Z}$, and these
are the only points where $\f(u)$ could have poles.
Consider the function
\beq
\Phi(u)=
\f(u)+\f(u+\frac{2\pi}{3})+\f(u+\frac{4\pi}{3}),\label{Phi1-def}
\eeq
which also satisfies \eqref{lame-pde} along with $\f(u)$. 
When $u=0$ the second and third terms in \eqref{Phi1-def} 
may have $n$-th order poles, however, they must cancel each other due to the 
last relation in \eqref{f-per}. Thus $\Phi(u)$ could have at most $(n-1)$-th 
order pole at $u=0$ (the first term in \eqref{Phi1-def} 
obviously vanishes due to \eqref{psi-cond}). As noted above
such pole is forbidden by eq.\eqref{lame-pde} and thus $\Phi(u)$ has 
the $(n+1)$-th order zero at $u=0$. 
Similarly, one concludes that $\Phi(u)$ vanishes at all points
 $u=k\pi/3+m\pi\tau$, \ with $k,m\in{\mathbb Z}$,
and, therefore, has at least
$12(n+1)$ zeroes in the periodicity parallelogram (of the periods
$2\pi$ and $2\pi\tau$) and no poles at all.
However, for $n\ge1$  this contradicts \eqref{f-per}, unless
$\Phi(u)\equiv0$, which is equivalent to \eqref{TQ1}. The special case
$n=0$ is considered in \cite{BMfuture}.

Now consider various limiting forms of the equation \eqref{lame-pde}.
When $q\to0$, with $u$ and $n$ kept fixed, the functions $\f_\pm(u)$
reduce to their analogs for the 6-vertex model
\beq
\f_\pm(u,q,n)=q{}^{\frac{3}{2}\,(d_\pm+\frac{1}{4})}\,
\f^{(6v)}_\pm(u,n)\,(1+O(q)),\qquad q\to0,
\eeq
where $d_\pm=(1\mp6)/36$ and eq.\eqref{lame-pde} becomes
\beq
\Big\{-\,\frac{d^2}{ds^2}+
\frac{n\,(n+1)}{\sin^2 s}-(d_\pm+{\ts\frac{1}{4}})\Big\}\f_\pm^{(6v)}(s/3,n)=0,
\eeq
which is simply related to eq.(13) of \cite{Str01a}. Taking now
$n\to\infty$ and $s\sim i\log n$ one recovers the eigenvalues of
the $\Q$-operators
of ref. \cite{BLZ97a} corresponding to $p=\pm1/6$
vacuum states in the $c=0$ CFT
\beq
Q_\pm^{(CFT)}(\t+\log t)=e^{\frac{\t}{2}}\,\lim_{n\to\infty}\,
\f_\pm^{(6v)}\left({\ts\frac{i}{3}}\log(8n/t)-{\ts\frac{i}{3}}\t\right),
\eeq
which are known \cite{Fen99} to satisfy the Bessel differential equation
\beq
\Big\{-\partial^2_\t+\partial_\t+\frac{1}{16}\,t^2\,
 e^{2\t} +d_\pm\Big\}Q_\pm^{(CFT)}(\t+\log t)=0.\label{Bessel}
\eeq

In the most interesting scaling limit \eqref{scaling}, the limiting
functions
\beq
\QQ_\pm(\t,t)=t^{1/4}\,e^{t^2/16}\,e^{\t/2}\, \lim_{n\to\infty}
\,n^{-1/4}\,\f_\pm(
\pi\tau/2-i\t/3,\,e^{i\pi\tau},\, n)\Big|_{\tau=\frac{2i}{3\pi}\log(8n/t)}
\label{Q-qft}\eeq
coincide with the eigenvalues \cite{Fen99} of the 
$\Q$-operators for special twisted vacuum states
in the massive sine-Gordon at the supersymmetric
point \cite{Zam94} (where the ground state energy of the model 
vanishes identically due to the supersymmetry).  
These eigenvalues 
satisfy the ``non-stationary Mathieu equation''
\eqref{mathieu-pde}. 
Let us show that this equation
completely determines the asymptotic expansion of these eigenvalues
at large $\theta$,
\beq
\log\QQ_\pm(\t,t)=-\frac{t}{4}e^\t+\sum_{k=0}^{\infty} \Bcal_\pm^{(k)}(t)\,
e^{-k\t},\qquad \t\to+\infty .\label{asympt}
\eeq
Consider the integral operator $\hat\Kcal(t)$ with the kernel
\beq
\Kcal(t|\t,\t')=\frac{1}{2\pi}\
\frac{e^{-u(\t)-u(\t')}}{1+e^{\t-\t'}},\qquad
u(\t)=\frac{t}{2}\,\cosh\t,\label{intop}
\eeq
which satisfies the following identity
\beq
\Big[-t\partial_t+{\hat{\Mcal}}_\t-{\hat\Mcal}_{\t'}\Big]\,
\Kcal(t|\t,\t')=\frac{1}{4\pi}\,t\,e^{-u(\t)-u(\t')+\t'},
\eeq
where $\hat\Mcal_\t$ denotes the differential operator in the
RHS of \eqref{mathieu-pde}.
Using this identity one can show that $\QQ_\pm(\t)$ satisfy the
linear integral equation discovered in \cite{Fen99}
\beq
\QQ_\pm(\t,t)=\Dcal_\pm(t)\,e^{-u(\t)}\ \mp\int_{-\infty}^{\infty}\
\Kcal(t|\t,\t')\ \QQ_\pm(\t',t)d\t'\label{int},
\eeq
provided that the functions $\Dcal_\pm(t)$ coincide with the Fredholm
determinants \cite{Zam94}
\beq
\Dcal_\pm(t)={\mathcal C}_\pm\,\det\big(\,1\pm\,\hat{\Kcal}(t)\,\big),
\eeq
where ${\mathcal C}_\pm$ are numerical constants.
Comparing \eqref{asympt} with \eqref{int} one concludes that
$\Bcal_\pm^{(0)}(t)=\log\Dcal_\pm(t)$. Then eq.\eqref{mathieu-pde} allows
to express all higher coefficients in \eqref{asympt} through the power
$\Bcal_\pm^{(0)}(t)$ and its derivatives. The few first coefficients
are shown in \eqref{asymp}. Finally note, that 
in the limit $t\to0$, \ $\t\sim-\log t$, 
\beq
\QQ_\pm(\t,t)=t^{d_\pm}\, Q^{(CFT)}_\pm\big(\t+\log t\big)\,
\big(1+O(t^{4/3})\big),\qquad 
t\to0,\qquad \t\sim-\log t,
\eeq
and the equation \eqref{mathieu-pde} reduces to \eqref{Bessel} as it,
of course, should\footnote{The normalization of
integral operator \eqref{intop} has been fixed from the comparison of
the $t\to0$ limit of the coefficients 
$\Bcal_\pm^{(1)}(t)$ with appropriate CFT
results of \cite{BLZ97a}. This normalization corresponds to
$\lambda=1/(4\pi)$ in eq.(3.1) of \cite{Zam94}.}.
 
We expect that our results can be readily extended 
to elliptic generalizations of the other special lattice models 
\cite{DTS04}, closely related to the $\eta=\pi/3$ six-vertex model. 
However, it would be more important to understand whether similar
considerations could be applied for the eight-vertex 
with an arbitrary value of $\eta$ and whether the general scheme of
correspondence between the $c<1$ CFT and ordinary differential
equations, developed in \cite{DT99b,BLZsdet,BLZ03}, can be extended to
the massive field theory (sine-Gordon model) with the use of partial
differential equations. The investigation of these questions is in progress.

\section*{Acknowledgments}

The authors thank I.M.~Krichever, S.L.~Lukyanov, S.M.~Sergeev, Yu.G.~Stroganov,
A.B.~Zamolodchikov
and Al.B.~Zamolodchikov for stimulating discussions and valuable remarks. 

\def\cprime{$'$}

\end{document}